\documentclass[aps,prc,twocolumn,floatfix]{revtex4}
\usepackage{epsfig}
\begin{document}
\title{J/$\psi$ suppression in a dense baryonic medium}
\author{Partha Pratim Bhaduri}
\affiliation{Variable Energy Cyclotron 
Centre, 1/AF Bidhan Nagar, Kolkata 700 064, India}            
\author{A. K. Chaudhuri}
\affiliation{Variable Energy Cyclotron 
Centre, 1/AF Bidhan Nagar, Kolkata 700 064, India} 
\author{Subhasis Chattopadhyay}
\affiliation{Variable Energy Cyclotron 
Centre, 1/AF Bidhan Nagar, Kolkata 700 064, India}            
\date{\today}
\begin{abstract}
We have examined the available latest SPS data on $J/\psi$ suppression in Pb+Pb and In+In collisions at 158 A GeV. Our employed model, with parameters fixed by $p$-$p$ and $p$-nucleus collisions, gives excellent description of NA50 and NA60 data on centrality dependence of $J/\psi$ suppression. The model is then applied to predict the centrality dependence of $J/\psi$ production in Au+Au collisions at FAIR energy domain. A much larger suppression  of  $J/\psi$ is predicted. In addition the possible effects of a baryon rich medium on $J/\psi$ production is also investigated. 
\end{abstract}

\maketitle

\section{Introduction}

J/$\psi$ suppression has long been recognized as an important signature for the occurrence of color deconfinement in nuclear collisions. If QGP is produced in the collision zone, the $c\bar{c}$ binding potential gets shielded by Debye screening of colored partons leading to the reduction in the J/$\psi$ yield~\cite{MS,Vogt}. However subsequent experimental investigations have revealed a considerable suppression of the charmonium production already present in proton-nucleus (p+A) collisions, where QGP or more generally formation of any secondary medium is not expected. In these reactions, the produced $c\bar{c}$ pair may interact with the cold nuclear medium of the target nucleus, hindering the formation of a bound state. To quantify the nuclear effects, data for different target nucleus are conventionally analyzed in the framework of the Glauber model~\cite{Gla59}, and the suppression is expressed in terms of an effective ``absorption'' cross-section $\sigma_{\rm J/\psi}^{eff}$. When studying J/$\psi$ suppression in nucleus-nucleus (A+A) collisions, a precise knowledge of nuclear effects is an essential pre-requisite to disentangle the genuine hot-medium effects.  With this approach, both NA50 and NA60 collaborations at SPS, observed significant anomalous suppression of J/$\psi$ yield, at 158 A GeV in Pb+Pb~\cite{Ale05} and In+In~\cite{Arn07} collisions respectively. Though the Pb+Pb data were found to be well explained by a variety of models~\cite{bl96,bl00,ca00,ch01,ch02,ch02b,Grand}, with or without incorporating deconfinement scenario, none of them satisfactorily reproduced the NA60 data. The theorized origin of the additional suppression thus remained unsolved and debated. However in both the measurements, the corresponding value of $\sigma_{\rm J/\psi}^{eff}$ was extracted from the data collected in p+A collisions at 400 GeV~\cite{NA50-400}. With the new measurements of charmonium in p+A collisions at 158 GeV~\cite{scompar}, where $\sigma_{\rm J/\psi}^{eff}$ turned out to be almost twice as large as that at 400 GeV, the NA60 experiment reported the relative charmonium yield in In+In collisions to be compatible within errors with absorption in cold nuclear matter; an anomalous suppression of about 25 - 30 $\%$ still remains visible in the most central Pb+Pb collisions. 

Till date no measurement exists on J/$\psi$ production in heavy-ion collisions below the top SPS energy, primarily due to their low production cross sections. The Compressed Baryonic Matter (CBM) experiment at FAIR~\cite{Peter}, in GSI, Germany is planning to perform for the first time, a detailed study of charmonium production in nuclear collisions, at beam energies $E_b = 10 - 40$ A GeV. The J/$\psi$ mesons produced at an early stage of the collisions might help in characterizing the confining status of the highly compressed baryonic medium, predicted to be produced in these collisions. In our earlier work~\cite{partha1}, we made an estimate of J/$\psi$ production cross sections in proton induced collisions at FAIR. For our study we employed and adapted the two component QCD based nuclear absorption model. Originally proposed by Qiu, Vary and Zhang~\cite{Qui}, the model treats conventional normal nuclear suppression in an unconventional manner. Model parameters were tuned by analyzing the available data for inclusive J/$\psi$ cross sections in high energy proton-proton and proton-nucleus collisions. Aim of the present paper, is to extend our studies further to calculate the centrality dependence of J/$\psi$ production in nuclear collisions at FAIR. In addition, the possible impact  of a high baryon density secondary  medium (confined/de-confined), expected to be produced as a result of the collision, will also be  investigated.

\section{Brief description of the model}

In our employed model, $J/\psi$ production in high energy hadronic collisions, is assumed to  be  a  factorisable two step   process,  (i)  formation  of  $c\bar{c}$  pair, which  is well accounted by perturbative QCD and (ii) formation of $J/\psi$ meson from the $c\bar{c}$ pair, which is non-perturbative in nature.  At the leading order in $\alpha_s$, the partonic contributions to $c\bar{c}$ production come from two subprocesses: quark annihilation ($q\bar{q} \rightarrow c\bar{c}$) and gluon fusion ($gg\rightarrow c\bar{c}$). With the K-factor accounting for effective higher order contributions, the single differential J/$\psi$ production cross section in collisions of hadrons  $h_1$ and $h_2$, at the center of mass energy $\sqrt{s}$ can be expressed as,
\begin{equation}
\label{diff}
\frac{d\sigma_{h_1h_2}^{J/\psi}}{dx_F} = K_{J/\psi}\int dQ^{2}\left(\frac{d\sigma_{h_1h_2}^{c\bar{c}}}{dQ^2dx_F}\right)\times F_{c\bar{c}
\rightarrow J/\psi}(q^2), 
\end{equation}
\noindent  where $Q^2 = q^2 +4 m_C^2$ with $m_C$ being the mass of the charm quark and $x_F$ is the Feynman scaling variable. $F_{c\bar{c} \rightarrow  J/\psi}(q^2)$ is the transition probability that a $c\bar{c}$ pair with relative momentum square $q^2$ evolve into a physical  $J/\psi$  meson, in hadronic collisions. Different parametric forms have been formulated for the transition probability following the existing models of color neutralization. Out of them two functional forms namely the Gaussian form ($F^{\rm (G)}(q^2)$) and power law form ($F^{\rm (P)}(q^2)$) respectively bearing the essential features of the  Color-Singlet~\cite{Singlet} and Color-Octet~\cite{Octet} models have been found to describe the $J/\psi$ production cross section data in p+A collisions reasonably well. In p+A collisions, charmonium production gets affected by the prevailing cold nuclear matter of the target nucleus. At the initial stage, nuclear modifications of the parton densities inside the target nucleus affect the perturbative $c\bar{c}$ pair production cross section. In our analysis, leading order MSTW2008~\cite{MSTW} set was used for free proton pdf and EPS09~\cite{EPS09} interface for the ratio $R_i(A,x,Q^2)$, that converts the free-proton distributions for each parton $i$, $f_i^{p}(x,Q^2)$, into nuclear ones, $f_i^A(x,Q^2)$. In nucleus-nucleus collisions, parton densities are modified both inside projectile and target nuclei. Depending on the collision geometry, either the halo or the core of the nuclei will be mainly involved, and the resulting shadowing effects will be more important in the core than in the periphery. Hence the shadowing factors have to be calculated for various centrality intervals. Assuming shadowing is proportional to the local nuclear density~\cite{Kle03, Eme99}, the spatial dependence is defined as:

\begin{equation}
R_{i,\rho}(A,x,Q^2,{\bf s},z)=1+N^A_\rho(R_i(A,x,Q^2)-1)\frac{\rho_A({\bf s},z)}{\rho_0},
\label{localpdf}
\end{equation}

\noindent where normalization $N^{A}_{\rho}$ is fixed to ensure that $(1/A)\int d{\bf s}dz R_{i,\rho}(A,x,Q^2,{\bf s},z)=R_i(A,x,Q^2)$. At large radii, $r(=\sqrt{(s^2+z^2)})>> R_A$ and $R_{i,\rho}\rightarrow 1 $, while at the nuclear centre, the modifications are larger than the average $R_i$.

\begin{figure} \vspace{-0.1truein}
\includegraphics[height=5.5cm,width=7.5cm]{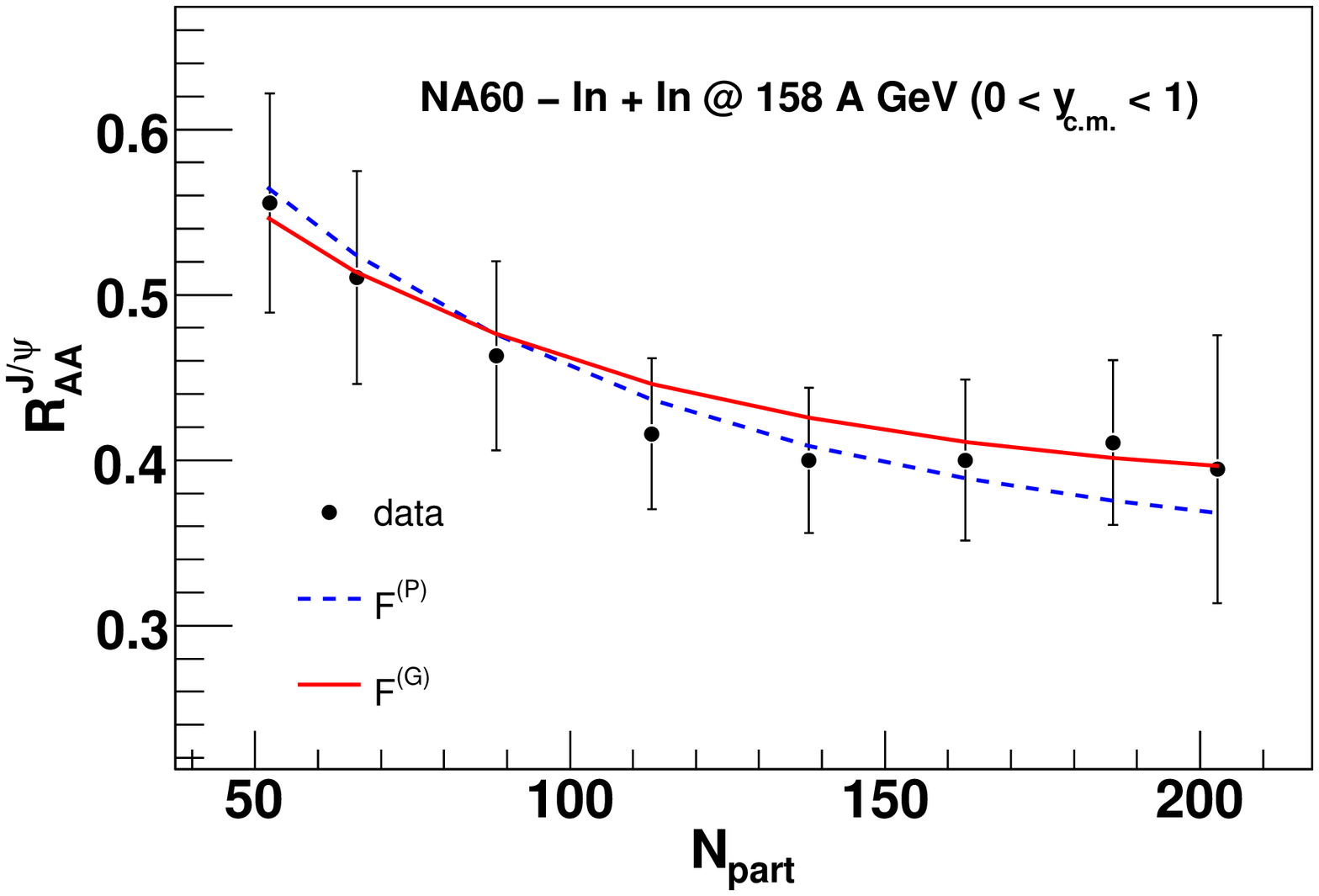}
\includegraphics[height=5.5cm,width=7.5cm]{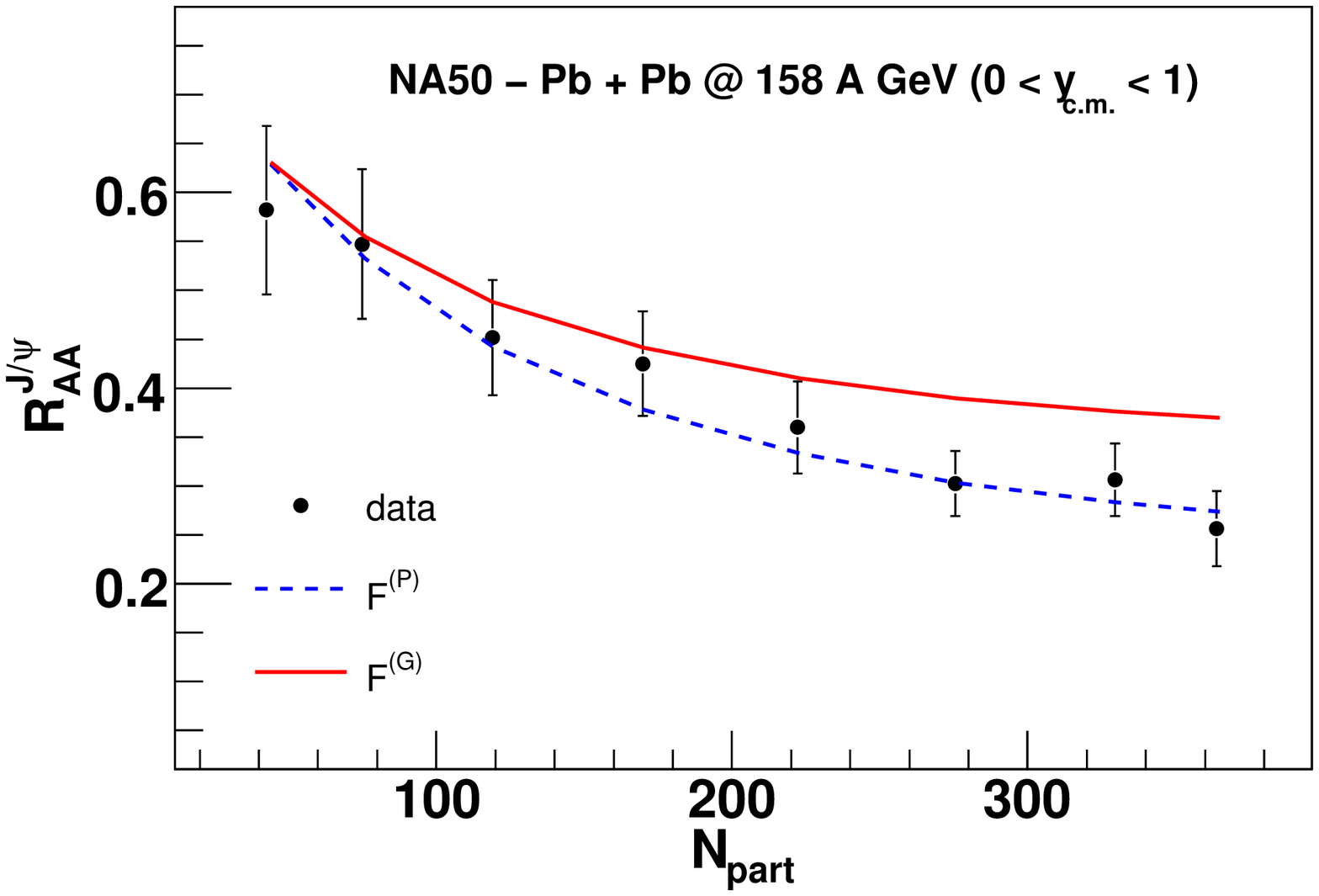}
\caption{\footnotesize Centrality dependence of J/$\psi$ production in In+In (top) and Pb+Pb (bottom) collisions measured at same energy ($E_b$ = 158 A GeV) and kinematic domain (0 $<y_{c.m.} <$ 1). Data are represented in terms of nuclear modification factor $R_{AA}$ plotted as a function of $N_{part}$ estimating the collision centrality. Error bars include both statistical and systematic uncertainties. Two different parametric forms of the transition function are used for generating the theoretical curves.} 
\label{fig1}
\end{figure}
Once produced, the nascent $c\bar{c}$ pairs interact with nuclear medium and gain relative square momentum at the rate of $\varepsilon^2$ per unit path length inside the nuclear matter. As a result, some of the $c\bar{c}$ pairs can gain enough  momentum to cross the threshold to become open charm mesons, leading  to the reduction in J/$\psi$ yield compared to the nucleon-nucleon collisions. For both parameterizations of transition probability, the corresponding values of $\varepsilon^2$, extracted from the analysis of p+A collision data~\cite{partha1}, exhibited non-trivial beam energy dependence. Lower be the beam energy, larger is the value of $\varepsilon^2$. In the present work we have used the previously found  $\varepsilon^2$ values. 

\section{Analysis of SPS data}

Let us now move forward to test the applicability of the model in describing the heavy-ion data on J/$\psi$ suppression at SPS. Fig.\ref{fig1} shows the variation of $R_{AA}^{J/\psi}$ as a function of $N_{part}$ for In+In and Pb+Pb collisions as calculated from our model in comparison with the available latest data~\cite{Arn11}. The In+In data points can be reasonably described within errors by both Gaussian ($F^{(G)}(q^2)$) as well as power law  ($F^{(P)}(q^2)$) forms of transition probability. In case of Pb+Pb collisions, $F^{(G)}(q^2)$ gives lower suppression than that observed in data. However $F^{(P)}(q^2)$ can fairly describe the data for all centralities and hence does not provide any additional room for any anomalous suppression mechanism to set in. For $F^{(G)}(q^2)$, the corresponding suppression is equivalent to that obtained in first order approximation of Glauber theory~\cite{partha1}. The corresponding value of $\epsilon^2$ was obtained  by analyzing the recent NA60 data for p+A collisions at 158 A GeV. Thus it can account for the In+In data but fails to generate enough suppression for Pb+Pb case. On the other hand due to threshold effect power law form generates a much stronger suppression for collisions involving heavy nuclei. As all the model parameters are constrained from the p+p and p+A data, in our present calculations, no free parameter is required to be tuned. The observed J/$\psi$ suppression in Pb+Pb collisions can be fully accounted for by the heavy quark re-scattering in the cold nuclear medium, without considering further suppression in the hot medium created in the later expansion stages. Earlier the model has also been found successful to describe the then available NA50 data on J/$\psi$ suppression in Pb+Pb collisions~\cite{ch02}. However in those studies shadowing corrections to nuclear parton densities were ignored and $E_T$ fluctuations had to be explicitly incorporated, through a tunable parameter, for better reproduction of the data at large $E_T$. 

\section{Predictions for FAIR energies}

Our ultimate goal is to estimate the J/$\psi$ yield in nuclear collisions at energies relevant to those available at FAIR. For this purpose, we will now use our model with the power law form of transition probability ($F^{(P)} (q^2)$) to calculate the centrality dependence of $R_{AA}^{J/\psi}$ for Au+Au reactions at a bombarding energy 25 A GeV. Previously predictions of J/$\psi$ survival probability at this energy were made within transport model calculations~\cite{HSD}. Nuclear effects were incorporated through conventional Glauber suppression scenario. For simulating the anomalous suppression two different scenarios namely 'QGP threshold melting' and 'hadronic co-mover absorption' were independently studied. For partonic scenario, a variant of the geometrical threshold model~\cite{bl96} was used with different melting energy densities for different charmonium states. For the hadronic dissociation, inelastic collisions with different mesons was considered. However in those calculations, the magnitude of the CNM effects at FAIR is possibly underestimated as the value of effective absorption cross section was taken from the p+A measurements at 400 GeV. Our present estimates predict a much larger nuclear suppression. Note that the degree of suppression induced by cold nuclear matter strongly depends on the passing time, $t_{d}=2R_{A}/\gamma$, of the two colliding nuclei, where $R_A$ is the nuclear radius and $\gamma$ is the Lorentz contraction factor. At SPS energy ($E_{c.m.} \simeq 17.3$ GeV) the collision time is about 1 fm/c and the magnitude of nuclear effects are large. At FAIR energies, the collision time ($t_d \simeq 3 fm/c$) is even much longer and the J/$\psi$ mesons during their evolution, will mostly encounter the (primary) nuclear medium rather than any secondary medium formed eventually due to the collision. Hence nuclear suppression will possibly play the most prominent role to govern the overall suppression pattern and we refrain ourselves to consider any probable additional suppression due to mesonic co-movers. However at FAIR energy regime, formation of highly compressed baryonic matter at low temperature is anticipated. Monte Carlo simulations~\cite{density} indicate the maximum baryon density in a central ($b=0$) Au + Au collision at FAIR energy to reach as high as $\rho_{B}=10\rho_{0}$. Thus the possible imprint of such a high density medium on J/$\psi$ production might be worth investigable. Since J/$\psi$ formation time ($\tau_{J/\psi}\simeq 0.5$ fm/c) is small compared to that required for the formation of any secondary medium, the highly compressed medium will most likely encounter the color neutral physical mesons rather than their precursors. The additional suppression induced by a confined baryon dense medium can be schematically expressed as:
\begin{equation}
S^{\rho_B}_{J/\psi}({\bf b,s}) =exp(-\int_{\tau_0}^{\tau_I}{d\tau\rho_B({\bf b,s},\tau)<v\sigma_{J/\psi-N}>})
\label{cosurvival} 
\end{equation}
In the above equation, $\sigma_{J/\psi-N} = 6.8$ mb~\cite{Oset} is the average inelastic cross section of the nucleons with the already formed J/$\psi$, $v \simeq 0.6$~\cite{Vogt} is J/$\psi$ velocity and $\rho_B({\bf b,s},\tau)$ is the net baryon density at proper time $\tau$ at the J/$\psi$'s position. Following~\cite{colb}, the spatial dependence of the net baryon number density is set with the transverse profile of the participant density, obtained in a Glauber model. $\tau_{0}$ and $\tau_{I}$ respectively denotes the medium formation time and the interaction time up to which J/$\psi$'s will continue interacting with the medium. Both of them will depend on the path length through the nucleus and can be obtained from~\cite{Vogt}. The evolution of baryon density with proper time $\tau$ can be followed from the equation for conservation of net baryonic current. If we neglect the transverse expansion (assuming that transverse expansion is slow and J/$\psi$ suppression occurs much before the transverse expansion sets in), we are left with, $\tau_{0}\rho_{B}(\tau_0) =\tau\rho_{B}(\tau)$.

\begin{figure} \vspace{-0.1truein}
\includegraphics[height=5.5cm,width=7.5cm]{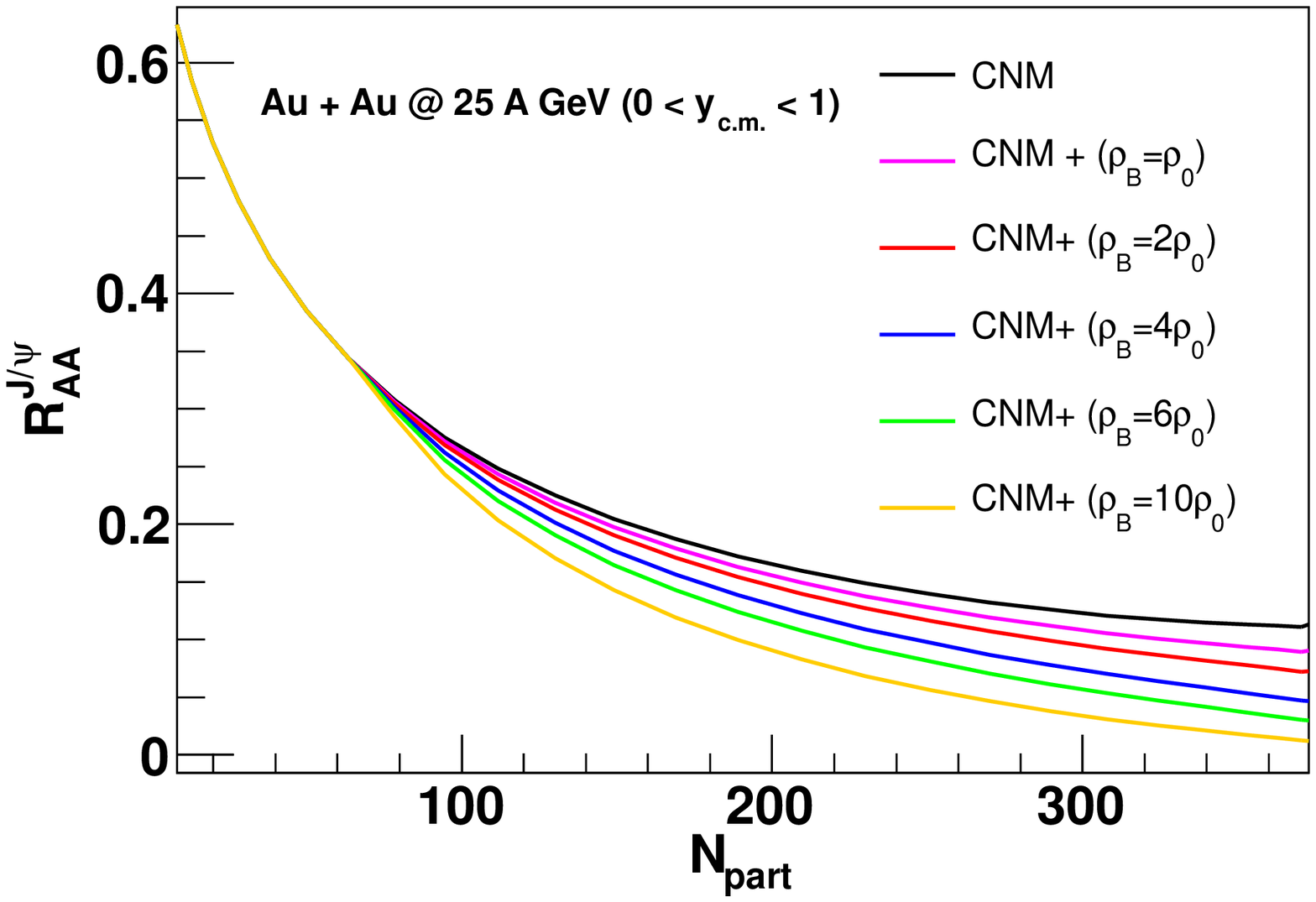}
\includegraphics[height=5.5cm,width=7.5cm]{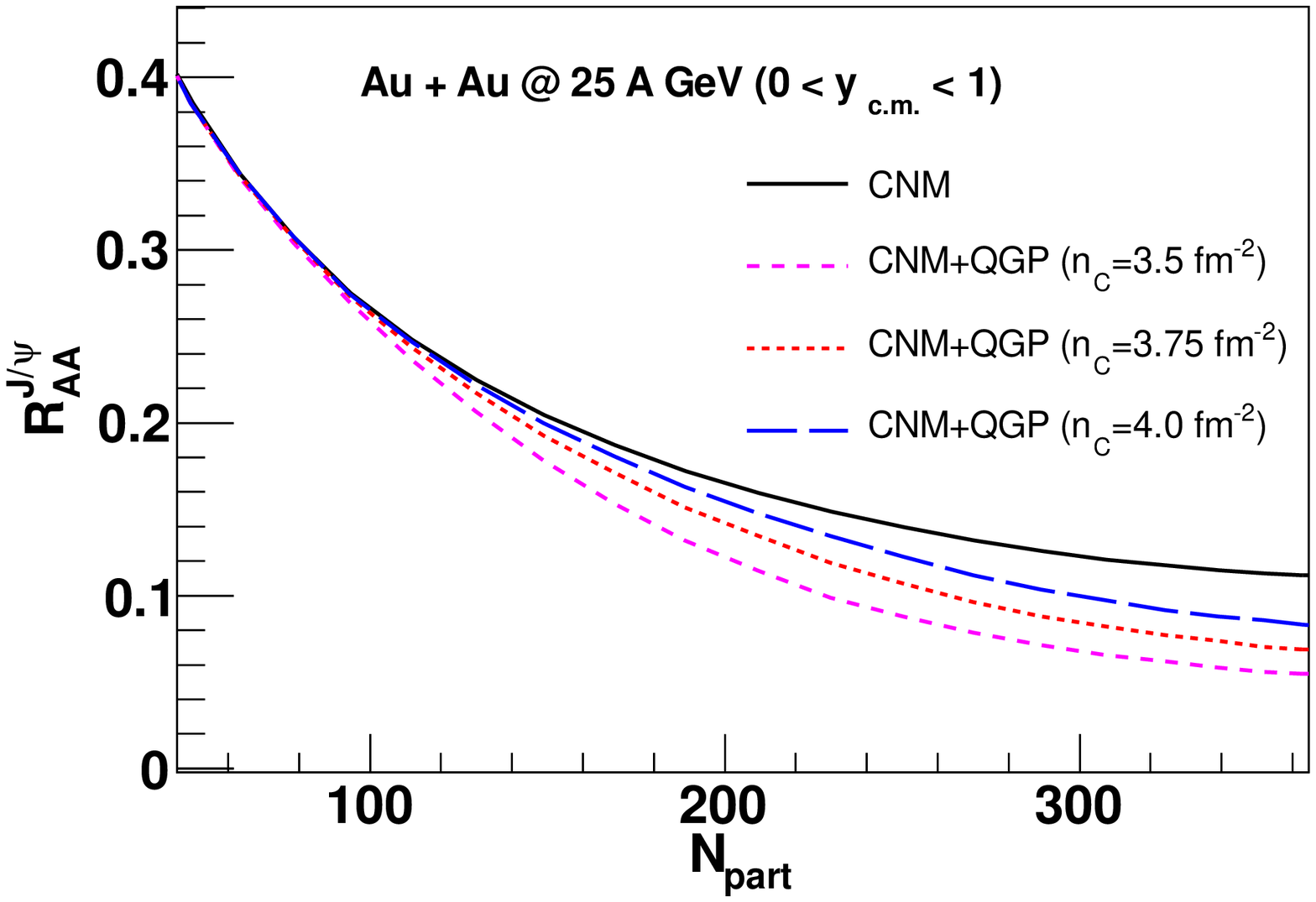}
\caption{\footnotesize Centrality dependence of J/$\psi$ suppression at 25 A GeV Au+Au collisions. In addition to nuclear effects, additional suppression due to high baryon density a) confined medium (top) and b) deconfined medium (bottom) are also shown.}
\label{fig2}
\end{figure}
 
The suppression pattern induced by a confined compressed baryonic medium is then shown in the top panel of Fig.\ref{fig2}. Calculations are performed for different peak densities varying from $\rho_0$ to $10\rho_0$. Higher be the density more violent is the suppression. If the maximum density of the produced medium is as high as $10\rho_0$, $R_{AA}^{J/\psi}$ approaches to zero and almost no J/$\psi$ will survive. However if such high density is achieved in the initial phase of the collision, deconfinement might set in resulting a phase governed by partonic degrees of freedom. In a partonic phase the  J/$\psi$ will interact differently with the medium. The interaction potential binding the $c$ and $\bar{c}$ together will be subject to Debye screening induced by the free color charges. To mimic the suppression pattern in a deconfined plasma, we follow the geometrical threshold model~\cite{bl96}, without considering the detailed microscopic dynamics. In this model the J/$\psi$ suppression function, at an impact parameter b, can be written as: 
\begin{equation}
S_{J/\psi}^{QGP} (b) = \int{d^2{\bf s}\Theta(n_c - n_p(b,{\bf s}))}
\label{qgp}
\end{equation}

\begin{figure} \vspace{-0.1truein}
\includegraphics[height=5.5cm,width=7.5cm]{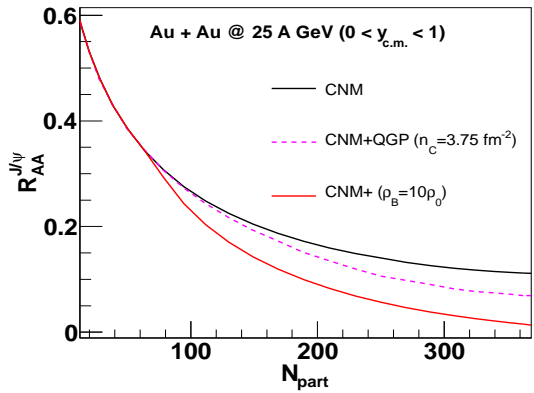}
\caption{\footnotesize Comparison of the two different scenarios of J/$\psi$ suppression in a compressed baryonic medium.}
\label{fig3}
\end{figure} 

The density $n_{p}(b,{\bf s})$ in the step function is proportional to the local energy density of the matter at position $(b,{\bf s})$. In the hot and dense part of the fireball where $n_{p}$ is larger than a critical/threshold value $n_c$, all the J/$\psi$ are absorbed in the medium and those outside this region only suffer normal suppression. The threshold density $n_c$ in this model is a parameter, generally fixed from the data. However it has been observed earlier that a critical density $n_c \simeq 3.6 - 3.7 fm^{-2}$ can reasonably describe both the data sets from SPS~\cite{bl00,ch01} and RHIC~\cite{akc}. $n_c$ can be thought of to be proportional to the threshold dissociation energy density ($\epsilon^{J/\psi}_d$) required for melting of J/$\psi$. If we assume a constant value of  critical energy density ($\epsilon_c \simeq 1 GeV/fm^3$), independent of baryon chemical potential $\mu_B$, required for deconfinement transition, then by analogy the threshold dissociation energy density ($\epsilon^{J/\psi}_d$) and consequently the critical participant density, $n_c$, can be assumed to be constant.  The right panel of Fig.\ref{fig2}, represents the behavior of $R_{AA}^{J/\psi}$ for three illustrative cases with three different critical densities. Smaller be the critical density, lower will be the energy density required for J/$\psi$ melting and more will be the suppression.  We put an end to this section by making a comparative study for these above two different mechanisms of anomalous suppression. For this purpose we consider two illustrative cases: a) confined baryonic medium with highest possible net baryon density ($\rho_B=10\rho_0$) and b) deconfined medium with approximately constant threshold energy (and hence participant) density. The results are shown in  Fig.\ref{fig3}. Two different mechanisms produce distinguishably different amount of suppressions. In a confined high baryon density medium, dissociation is more severe compared to that in QGP phase. Thus measurement of J/$\psi$ production in nuclear collisions at FAIR might also furnish valuable information about the phase structure and the relevant degrees of freedom in such a high baryon density environment never observed before. 

\section{Summary}
In summary, we have estimated the J/$\psi$ production and its possible interactions in a high baryon density medium anticipated in low energy nuclear collisions at FAIR. Our model satisfactorily describes the J/$\psi$ suppression data in heavy-ion collisions at SPS, with model parameters being fixed from p+p and p+A data. At FAIR, exogamous production in both the partonic and hadronic phase is expected to be small and the primordial production  will dominate the overall J/$\psi$ yield. Consequently such measurements will offer us the golden opportunity to exactly trace out the possible suppression pattern which will not get masked by the subsequent regeneration. Moreover at low energies, collision time is much longer and the lifetime of the produced medium is much shorter and nuclear effects start playing a dominant role in deciding the observed charmonium yield. Even at SPS, the magnitude of the nuclear effects, in our employed framework, are substantially large to fully account for the observed J/$\psi$ suppression in Pb+Pb collisions. At FAIR effects of the cold nuclear matter will be further amplified leading to a strong reduction of the J/$\psi$ yield in most central collisions. The fully formed J/$\psi$ mesons surviving the nuclear dissociation can subsequently interact with produced high density medium and undergo further suppression. The degree as well as the mechanism of this additional suppression depends on the net baryon density achieved in the collision and the confining status of the medium.

\end{document}